\documentclass[a4paper]{jpconf}
\usepackage{graphicx}
\usepackage{bm}
\begin{document}
\title{Shell model study of first-forbidden beta decay around $^{208}$Pb}

\author{P. C. Srivastava, S. Sharma, A. Kumar and P. Choudhary}

\address{Department of Physics, Indian Institute of Technology Roorkee, Roorkee 247667, India}

\ead{praveen.srivastava@ph.iitr.ac.in}

\begin{abstract}
In the present work, we report a systematic theoretical study of the $\log ft$ values for the forbidden $\beta^-$ decay transitions in the $^{208}$Pb region. For this, we have considered $^{206}$Hg $\rightarrow$ $^{206}$Tl,  $^{208}$Hg $\rightarrow$ $^{208}$Tl,  
$^{206}$Tl $\rightarrow$ $^{206}$Pb and $^{208}$Tl $\rightarrow$ $^{208}$Pb transitions. We have performed shell model calculations using KHH7B interaction 
in valence shell 58-114 for protons and 100-164 for neutrons by considering ${\it 1p-1h}$ excitations for both protons and neutrons simultaneously  for daughter nuclei. 
This study presents the first shell model results of $\beta^-$-decay corresponding to the recent experimental data.

\end{abstract}

\section{Introduction}

The neutron-rich nuclei around doubly magic nucleus $^{208}$Pb undergo $r$-process, i.e., rapid neutron capture nucleosynthesis, which can be seen from the third peak of abundance pattern in solar's elemental composition. In order to model these abundance patterns and determine the time scale of explosive stellar processes, the study of radioactive decay of the nuclei in this region is very important. Since reproducing these elements is quite challenging experimentally, thus, studying the beta decay properties of these nuclei using theoretical models is highly desirable. Several experimental studies have been conducted in the past to evaluate the $\log ft$ values in the region below and above $^{208}$Pb \cite{Carroll,Berry,Brunet}.
The allowed transitions compete with the first-forbidden $\beta$ decay transitions which is recently reported in \cite{Carroll} corresponding to  $^{208}$Hg $\rightarrow$ $^{208}$Tl transition  from the newly performed experiment at CERN-ISOLDE. In this work, the $\log ft$ values associated with three excited states of $^{208}$Tl were reported for the first time. In the present study, apart from these transitions, we have also computed the $\log ft$ values corresponding to
$^{206}$Hg $\rightarrow$ $^{206}$Tl,  
$^{206}$Tl $\rightarrow$ $^{206}$Pb and $^{208}$Tl $\rightarrow$ $^{208}$Pb transitions \cite{nndc}.

\section{ Formalism for beta decay}

In the beta decay formalism, the partial half-life of the transition is given by

\begin{eqnarray}\label{hf1}
t_{1/2}=\frac{\kappa}{\tilde{C}},
\end{eqnarray}

where constant $\kappa$  is  6289 s \cite{Patrignami} and the term $\tilde{C}$ denotes the dimensionless integrated shape function and it can be written in the form

\begin{eqnarray} \label{tc}
\tilde{C}=\int_1^{w_0}(w_0-w_e)^2(w_e^2-1)^{1/2}w_eC(w_e)F_0(Z,w_e)dw_e.
\end{eqnarray}

The shape factor $C(w_e)$ for the allowed transition is given in the following form:
\begin{equation}
C(w_e)  = \frac{g_{A}^{2}}{2J_{i}+1}|\mathcal{M}_{GT}|^2,
\end{equation}

where, the term $\mathcal{M}_{GT}$ stands for the Gamow-Teller nuclear matrix element (NME). In Gamow-Teller transition, 
$C(w_e)$ is known as reduced transition probability [$B(GT)$]. The  angular momentum of initial state is denoted by $J_{i}$ and the term $g_{A}$ stands for the axial-vector coupling constant. 
In this case, the partial half-life and the phase-space factor, i.e., Fermi integral ($f_0$) are multiplied to provide the reduced half-life ($ft$),  where $f_0$ is defined as
\begin{equation}
	f_{0}= \int_1^{w_0}(w_0-w_e)^2(w_e^2-1)^{1/2}w_eF_0(Z,w_e)dw_e.
\end{equation}
Generally, the $ft$ values are expressed as $log ft$ values, which are given as

\begin{equation}
	{\rm log}~ {\it ft}  = {\rm log}_{10} (f_{0}t_{1/2}[s]).
\end{equation}

The expression for the shape factor, i.e., $C(w_e)$ for forbidden transition and other details can be obtained from earlier works \cite{mst2006,anil1,anil2,Sharma,anil5}.

\begin{figure}
\includegraphics[scale=0.6]{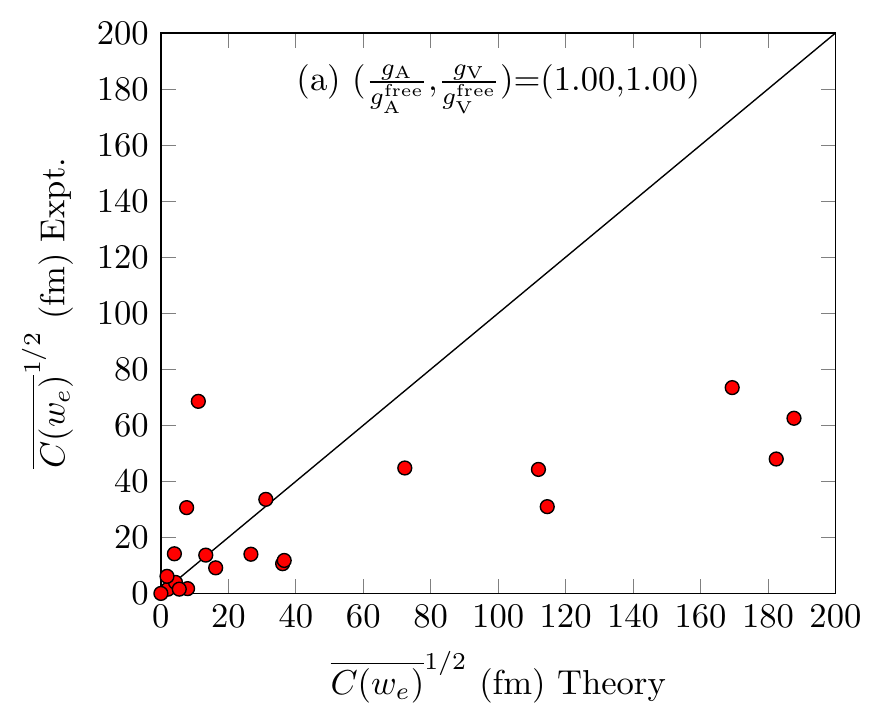}
\includegraphics[scale=0.6]{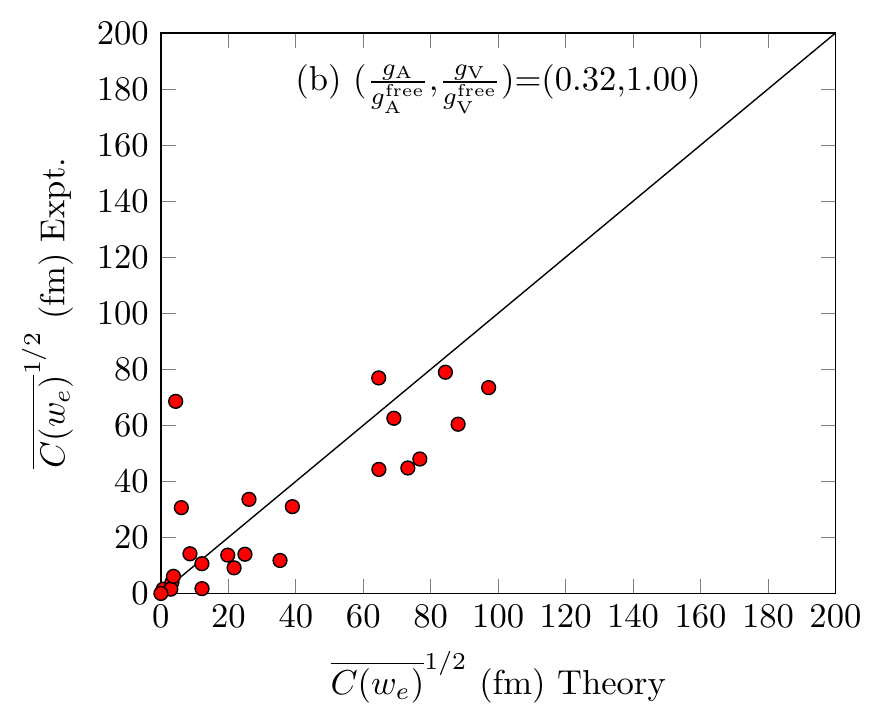}
\includegraphics[scale=0.6]{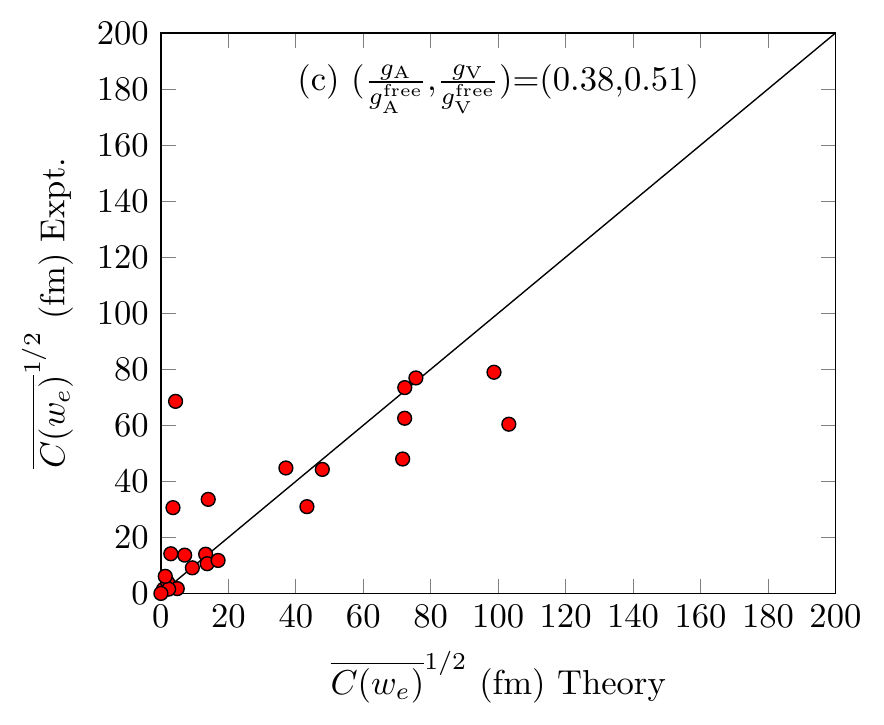}
\caption{\label{fig1}The calculated and experimental average shape factor for first-forbidden and allowed 
transitions corresponding to axial-vector and vector coupling constants by using different values of weak coupling constants.}
\end{figure}

The NMEs associated with both allowed and forbidden beta decay transitions contain all of the nuclear-structure information and are given in the form

\begin{equation}
^{V/A}\mathcal{M}_{KLS}^{(N)}(pn)(k_e,m,n,\rho)   =\frac{\sqrt{4\pi}}{\widehat{J}_i}
\sum_{pn} \, ^{V/A}m_{KLS}^{(N)}(pn)(k_e,m,n,\rho)(\Psi_f\parallel [c_p^{\dagger}\tilde{c}_n]_K\parallel \Psi_i),
\label{eq:ME}
\end{equation}
where ${\widehat{J}_i}=\sqrt{2J_i+1}$ and the summation runs over the proton $(p)$ and neutron $(n)$ single-particle states.  This equation has two parts: 
the term ${^{V/A}m_{KLS}^{(N)}}(pn)(k_e,m,n,\rho)$ is nuclear model independent term and known as single-particle matrix elements (SPMEs) and $(\Psi_f|| [c_p^{\dagger}\tilde{c}_n]_K || \Psi_i)$ are the one-body transition densities (OBTDs) with $(\Psi_i)$ and $(\Psi_f)$ being the initial and final state wave functions, respectively. This OBTD is obtained from the shell-model diagonalization using NUSHELLX \cite{nushellx} code.

\begin{table}
\caption{\label{tabone} Comparison between shell-model calculated and experimentally measured $\log ft$ values. We have performed calculations with 
two sets of quenching factors for the weak coupling constants $g_{\rm V}$ and $g_{\rm A}$. The quenching factor values for the set I: ($g_{\rm A}/g_{\rm A}^{\rm free}$,$g_{\rm V}/g_{\rm V}^{\rm free}$)=(0.32,1.00) and set II: ($g_{\rm A}/g_{\rm A}^{\rm free}$,$g_{\rm V}/g_{\rm V}^{\rm free}$)=(0.38,0.51) with mesonic enhancement factor as $\epsilon_{MEC}$ = 2.01 have been taken. Here, FNU stands for forbidden non-unique, while FU stands for forbidden unique transitions.}  

\begin{center}

\begin{tabular}{|cc|c|cc|ccc|c|c|}
\hline  
\multicolumn{2}{|c|}{Transition}  &Decay mode&\multicolumn{2}{c|}{Energy (keV) } & \multicolumn{3}{c|}{$\log ft$}  \\
\cline{1-2}
\cline{4-5}
\cline{6-8}  
\hline                          
Initial &Final &  &Expt. & SM   & Expt.    &  SM$_{\rm {Set I}}$   & SM$_{\rm {Set II}}$ \\
\hline 
 
 $^{206}$Hg($0^+) $ & $^{206}$Tl($0_1^-$)  & 1st FNU   &  0.0 &0.0 &  5.41(6) & 5.082& 4.945  \cr
  & $^{206}$Tl($1_1^-$)  & 1st FNU   &  304.896(6)&312 &  5.24(10) & 4.997 & 5.254  \cr
  & $^{206}$Tl($1_2^-$)  & 1st FNU   &  649.42(4)   & 729  &  5.67(8)  & 5.243 & 5.834\cr 

  \hline

 $^{206}$Tl($0^-) $ & $^{206}$Pb($0_1^+$)  & 1st FNU   &  0.0   & 0.0  &  5.1775(13) & 5.120& 4.983  \cr

  & $^{206}$Pb($2_1^+$)  & 1st FU   &  803.049(25)   & 839  &  8.60(3) & 9.303  & 9.174 \cr  
  & $^{206}$Pb($0_2^+$)  & 1st FNU   &  1166.4(5)   &  1226 &  5.99(6) & 5.836 & 5.699  \cr
  
  \hline
  
  $^{208}$Hg($0^+) $ & $^{208}$Tl(($0_1^-$))  & 1st FNU   &  1806.7(3)   & 4290  &  5.2(1)  &5.352 & 5.215 \cr
  & $^{208}$Tl(($1_1^-$))  & 1st FNU   &  1960.0(11)   &  2332 &  6.0(2)  & 7.406  & 7.862\cr
  & $^{208}$Tl(($1_2^-$))  & 1st FNU   &  2119.0(4)   & 3213  &  5.3(1)  & 7.689  & 7.695\cr

    \hline
  
 $^{208}$Tl($5^+) $ & $^{208}$Pb($5_1^-$)  & 1st FNU   &  3197.717(11)   & 3231  &  5.61(1) & 5.202 & 5.261\cr
   & $^{208}$Pb($4_1^-$)  & 1st FNU  & 3475.088(11)   &  3425 &  5.68(1) & 5.351 & 5.612 \cr
   & $^{208}$Pb($5_2^-$)  & 1st FNU   &  3708.41(7)   &  3536 &  5.38(1)  & 5.293 & 5.254\cr
    & $^{208}$Pb($6_1^-$)  & 1st FNU  & 3919.78(10)   & 3880  &  6.68(4) & 6.179 &  6.726\cr
      & $^{208}$Pb($4_2^-$)  & 1st FNU  & 3946.42(20)   &  3780 &  7.78(3) & 7.943 & 8.370 \cr
  & $^{208}$Pb($5_3^-$)  & 1st FNU   &  3960.93(7)   & 3879  &  5.92(1)  & 6.138 & 6.678\cr 
    & $^{208}$Pb($4_3^-$)  & 1st FNU  & 3995.6(5)   & 3972  &  8.5(2) & 6.800 & 7.598 \cr
    & $^{208}$Pb($5_4^-$)  & 1st FNU  &  4125.28(17)   &  3996 &  6.92(6)  & 6.802 & 6.697\cr 
    & $^{208}$Pb($5_5^-$)  & 1st FNU &  4180.38(17)   & 4056  &  6.70(2)  & 6.378  & 7.275\cr 
      & $^{208}$Pb($4_4^-$)  & 1st FNU  & 4262.0(7)   & 4106  &  8.6(2) & 8.036 & 8.252 \cr
      & $^{208}$Pb($5_6^-$)  & 1st FNU  & 4296.28(20)   &  4152 &  6.83(9) & 5.876 & 6.513 \cr 
      & $^{208}$Pb($4_1^+$)  & Allowed  & 4323.4(4)   & 4708  &  8.1(2) & 9.894 & 9.757 \cr
    & $^{208}$Pb($4_5^-$)  & 1st FNU  & 4358.44(17)   & 4230  &  7.05(5) & 6.299 & 7.032 \cr
       & $^{208}$Pb($6_2^-$)  & 1st FNU  & 4382.9(3)   & 3993  &  7.4(2) & 7.830 &  8.746\cr
          & $^{208}$Pb($6_3^-$)  & 1st FNU  & 4480.5(3)   &  4216 &  6.67(5) & 7.101 &  8.035\cr
 \hline         
       
\end{tabular}
\end{center}
\end{table}

\section{Results and discussions}
In the present study, we have used the KHH7B interaction \cite{Hosaka,Kuo,Warburton} to carry out systematic large-scale shell-model computations corresponding to the parent and daughter nuclei in the model space spanned by proton number ($58 \leq Z \leq 114$), and neutron number ($100 \leq N \leq 164$) around the $^{208}$Pb nucleus.
This model space is spanned by proton orbitals $1d_{5/2,3/2}$, $0h_{11/2,9/2}$, $2s_{1/2}$, $1f_{7/2}$  and $0i_{13/2}$ around $Z=82$, and neutron orbitals $0i_{13/2,11/2}$, $2p_{3/2,1/2}$, $1f_{5/2}$, $1g_{9/2}$, and $0j_{15/2}$ around $N = 126$. Then, the OBTDs that are required for the calculations of the NMEs have been computed. In the present work, the $1p-1h$ truncation is employed for both protons and neutrons across the $^{208}$Pb for the daughter nuclei. Additionally, in $\Delta J=0^-$ transitions, the recoil-axial matrix element, i.e., $\gamma_5=\bm{\sigma} \cdot \bm{p_e}$  is improved with the help of mesonic enhancement factor, i.e., $\epsilon_{\rm{MEC}}$ on top of impulse approximation value. In this work, $\epsilon_{\rm{MEC}}=2.01$ is used which is taken from the Ref. \cite{warburton} for $^{208}$Pb region.

The comparison between theoretical and experimental average shape factors is shown in Fig. \ref{fig1}.
First, the $\log ft$ values with the bare values of the weak coupling constants, i.e., without the inclusion of the quenching factor have been calculated; in the second set quenching factor 
0.32 is included, which we obtained with chi-squared fitting method, such that $g_A$= $q g_A^{free}$ = 0.41 and $g_V$ = 1.00; finally the another set of weak coupling constants is included which is taken  from Ref \cite{Zhi}. 
In our calculations for $\log ft$ values, we have taken two sets of weak coupling constants, set I ($g_A/g_A^{free}$, $g_V/g_V^{free}$) = (0.32, 1.00), and set II with  ($g_A/g_A^{free}$, $g_V/g_V^{free}$) = (0.38, 0.51).
The calculated $\log ft$ values for the forbidden $\beta^-$ decay corresponding to  $^{206}$Hg $\rightarrow$ $^{206}$Tl,  $^{208}$Hg $\rightarrow$ $^{208}$Tl,  $^{206}$Tl $\rightarrow$ $^{206}$Pb and $^{208}$Tl $\rightarrow$ $^{208}$Pb transitions are reported in the Table  \ref{tabone}. 
In the case of $^{206}$Hg($0^+$) to  $^{206}$Tl($1_1^-$) transition, the experimental $\log ft$ value is 5.24(10) although it is 4.997 from set I and 5.254 from set II.
Further, on considering $^{206}$Tl($0^-$) to  $^{206}$Pb($0_1^+$) and $^{206}$Tl($0^-$) to  $^{206}$Pb($0_2^+$) transitions, our calculated results for set I are very close to the experimental $\log ft$ values. In the case of the first forbidden unique decay corresponding to 
$^{206}$Tl($0^-$) to  $^{206}$Pb($2_1^+$), our calculated values are slightly larger than the experimental data.
For the $^{208}$Hg $\rightarrow$ $^{208}$Tl transition, we have taken the experimental data from Ref. \cite{Carroll}. For 
$^{208}$Hg($0^+$) to  $^{208}$Tl($0_1^-$) transition, both sets give  excellent agreement with the experimental data. For 
$^{208}$Hg($0^+$) to  $^{208}$Tl($1_1^-$) transition, both sets provide larger values compared to the experimental value. Further, the deviation between theoretical and experimental values is increased for $^{208}$Hg($0^+$) to  $^{208}$Tl($1_2^-$) transition. Our results might be improved if we take extended model space such as KHM3Y interaction having 24 orbitals \cite{Brown}.
For $^{208}$Tl $\rightarrow$ $^{208}$Pb transition, the $\log ft$ values associated with 15 transitions have been calculated. Overall, there is a good agreement between our shell-model calculated values and the experimental data. However, the beta decay associated with higher excited states needs more particle-hole excitation ($3p-3h$), which we will consider in our future study.
\section{Summary and conclusions}
In this work, the $\log ft$ values for the allowed and forbidden $\beta^-$ decay transitions in the $^{208}$Pb region are reported. For this, we have considered $^{206}$Hg $\rightarrow$ $^{206}$Tl,  $^{208}$Hg $\rightarrow$ $^{208}$Tl,  
$^{206}$Tl $\rightarrow$ $^{206}$Pb, and $^{208}$Tl $\rightarrow$ $^{208}$Pb transitions. The beta decay properties using the large scale shell-model calculations agree very well with the corresponding experimental data.  We need extended model space in the cases where  $\log ft$ values are not close to the experimental data.
\vspace{-2mm}
\section*{Acknowledgment}

We acknowledge a research grant from SERB (India). SS would like to thank CSIR for financially supporting 
her Ph.D. thesis work.

\end{document}